\documentclass[sigchi,authorversion]{acmart}

\newcommand{\hl}[1]{#1}

\PassOptionsToPackage{hyphens}{url}
\usepackage{url}
\usepackage{xurl}
\usepackage[normalem]{ulem}
\newcommand\redsout{\bgroup\markoverwith{\textcolor{red}{\rule[0.5ex]{2pt}{0.4pt}}}\ULon}
\usepackage{xcolor}

\usepackage[table]{xcolor}
\usepackage{array}
\newcolumntype{L}[1]{>{\raggedright\arraybackslash}p{#1}}

\AtBeginDocument{%
  }

\copyrightyear{2026}
\acmYear{2026}
\setcopyright{cc}
\setcctype{by}
\acmConference[CHI '26]{Proceedings of the 2026 CHI Conference on Human Factors in Computing Systems}{April 13--17, 2026}{Barcelona, Spain}
\acmBooktitle{Proceedings of the 2026 CHI Conference on Human Factors in Computing Systems (CHI '26), April 13--17, 2026, Barcelona, Spain}
\acmPrice{}
\acmDOI{10.1145/3772318.3790640}
\acmISBN{979-8-4007-2278-3/2026/04}

\acmSubmissionID{8154}

\begin{document}
\title{Animated Public Furniture as an Interaction Mediator: Engaging Passersby In-the-Wild with Robotic Benches}

\author{Xinyan Yu}
\email{xinyan.yu@sydney.edu.au}
\orcid{0000-0001-8299-3381}
\affiliation{Design Lab, Sydney School of Architecture, Design and Planning
  \institution{The University of Sydney}
  \city{Sydney}
  \state{NSW}
  \country{Australia}
}
\author{Marius Hoggenmueller}
\email{marius.hoggenmueller@sydney.edu.au}
\orcid{0000-0002-8893-5729}
\affiliation{Design Lab, Sydney School of Architecture, Design and Planning
  \institution{The University of Sydney} 
  \city{Sydney}
  \state{NSW}
  \country{Australia}
}

\author{Xin Lu}
\email{xin.lu@sydney.edu.au}
\orcid{0000-0002-3754-9182}
\affiliation{School of Electrical and Computer Engineering
  \institution{The University of Sydney} 
  \city{Sydney}
  \state{NSW}
  \country{Australia}
}

\author{Ozan Balci}
\email{ozan.balci@kuleuven.be}
\orcid{0000-0002-9265-2262}
\affiliation{
  \institution{Research[x]Design -- Department of Architecture}
  \institution{KU Leuven}
  \city{Leuven}
  \country{Belgium}
}

\author{Martin Tomitsch}
\email{Martin.Tomitsch@uts.edu.au}
\orcid{0000-0003-1998-2975}
\affiliation{Transdisciplinary School,
  \institution{University of Technology Sydney}
  \city{Sydney}
  \state{NSW}
  \country{Australia}
}

\author{Andrew Vande Moere}
\authornote{These authors contributed equally to this work.}
\email{andrew.vandemoere@kuleuven.be}
\orcid{0000-0002-0085-4941}
\affiliation{
  \institution{Research[x]Design -- Department of Architecture}
  \institution{KU Leuven}
  \city{Leuven}
  \country{Belgium}
}

\author{Alex Binh Vinh Duc Nguyen}
\authornotemark[1]
\email{alex.nguyen@kuleuven.be}
\orcid{0000-0001-5026-474X}
\affiliation{
  \institution{Research[x]Design -- Department of Architecture}
  \institution{KU Leuven}
  \city{Leuven}
  \country{Belgium}
}
\affiliation{
  \institution{Faculty of Design Sciences}
  \institution{University of Antwerp}
  \city{Antwerp}
  \country{Belgium}
}

\renewcommand{\shortauthors}{Yu et al.}

\begin{abstract}
Urban HCI investigates how digital technologies shape human behaviour within the social, spatial, temporal dynamics of public space. Meanwhile, robotic furniture research demonstrates how the purposeful animation of mundane utilitarian elements can influence human behaviour in everyday contexts. Taken together, these strands highlight an untapped opportunity to investigate how animated public furniture could mediate social interaction in urban environments. In this paper, we present the design process and in-the-wild study of mobile robotic benches that reconfigure with a semi-outdoor public space. Our findings show that the gestural performance of the benches manifested three affordances perceived by passersby, they activated engagement as robots, redistributed engagement as spatial elements, and settled engagement as infrastructure. We proposed an Affordance Transition Model (ATM) describing how robotic furniture could proactively facilitate transition between these affordances to engage passersby. Our study bridges robotic furniture and urban HCI to activate human experience with the built environment purposefully.
\end{abstract}




\maketitle

\section{Introduction}

Public space, as noted by \citet{carr1992public}, serves human needs in three key ways: passive engagement with the environment, active engagement through interaction or stimulation, and the excitement of discovery. Within HCI, previous research has sought to activate these dimensions by integrating interactive technologies to the urban environment~\cite{Memarovic2012DiscoveryPublicSpace, Fischer2012UrbanHCI}, ranging from media building facades~\cite{Nguyen2023Xylophonic} to situated public displays~\cite{Biedermann2021CriticalReview,muller2012Lookingglass}. Studies on these interventions have shown that public engagement with urban-integrated interactive technologies is highly shaped by the surrounding context, including spatial setup~\cite{Fischer2012UrbanHCI} and social dynamics~\cite{Wouters2016HoneyPot, Andrew2012Context}. At the same time, these studies also demonstrate how interface affordances and associations impact spatial arrangements in ways that condition engagement, for example, determining whether an intervention succeeds in creating a space for social interaction~\cite{Fischer2012UrbanHCI}.



Alongside these digital interventions, the emerging domain of robotic furniture has similarly sought to foster engagement by animating physical elements of the built environment, actively shaping social interactions and spatial engagement in ways more deeply embedded in the surrounding context~\cite{ju2009approachability,Sirkin2015Ottoman}. Pioneering design studies in this domain have explored how furniture robots can convey autonomous intentions by performing dynamic gestures inspired by the way people normally utilise static furniture. For example, a door indicated approachability through opening and closing ~\cite{ju2009approachability}, a chair invited people to sit by moving backward~\cite{LC2024SitonMe,Agnihotri2019PersuasiveChairBots}, or a footrest withdrew its support by retreating away~\cite{Sirkin2015Ottoman}. Since then, researchers have extended this functionality-inspired approach to employ more diverse types of gesturing, such as a chair using \hl{emblematic} “side-to-side” gesture to negotiate walking turns with passersby~\cite{Knight2017ChairBotGesture}, a pair of stools using deictic “rotating towards each other” gesture to nudge social interaction between seated occupants~\cite{Guo2023TalkwithStrangers}, or a pair of partitions using architectonic “opening up” gesture to nudge eye contact and thus foster connection between co-workers~\cite{Ozan2025Meetings}. To date, in-the-wild deployment of robotic furniture has primarily been investigated in indoor settings, from private homes~\cite{Di2011RoboticTable}, to semi-private offices~\cite{Nguyen2024Adaptive}, where occupants inhabit the environment and become familiar with the robotic furniture over time. This raises the question of how the capabilities of indoor robotic furniture, both in conveying immediate affordances and in shaping socio-spatial dynamics, might be transferred into outdoor public spaces that are also populated with a wide variety of urban furniture.

Recent work by ~\citet{Brown2024Public} has begun to extend this line of inquiry by examining the spontaneous interactions between passers-by with robotic trashcans as an animated form of urban furniture. Particularly, their familiar affordances that are strongly tied to the urban context allowed them to blend seamlessly into municipal systems, eliciting natural and sequential interactions with passers-by. \hl{While \citet{Brown2024Public}'s investigation centre the interaction on the robotic artefact itself, our work shifts the focus from the robot to the situated interactions it mediates, asking: \textit{How can robotic furniture mediate people's socio-spatial engagement in public space, and how can we design for such mediation?}} Rather than layering digital interfaces onto the urban fabric, we explore the untapped potential to transform these mundane, familiar infrastructures into interactive artefacts that activate public engagement\hl{, not only through social interactions among people but also by mediating their engagement with the surrounding spatial context.}


Following a research-through-design approach~\cite{Zimmerman2007RtD}, we first conducted multiple iterative design explorations \hl{both to understand the potential roles robotic public furniture may play in mediating socio-spatial engagement and to explore how gestural strategies can be designed to enact these potentials.} This including field observations, a lab-based design workshop, and an in-situ trial. Through this design process, we achieved a \textit{gestural sequence} for the benches to grab the attention of passersby, invite them to linger, guide them towards an often-overlooked artistic sculpture in the library arcade, and nudge them to sit down and observe it.
To evaluate this gestural sequence in practice, we conducted an in-the-wild study using the Wizard-of-Oz approach~\cite{dahlback1993wizard}, running for 26 hours across five days, during which we observed 98 passerby engagements and conducted 55 \hl{brief in-situ} semi-structured interviews. 
Our findings unpack how individual passersby interpreted the gestural sequence performed by the robotic benches, which in turn shaped their direct interaction with the benches, mediated their interaction with the surrounding spaces, motivated their engagement with often-overlooked spatial elements, and nudged their social interactions both within and beyond their groups. 
We discuss how these diverse ways of engagement were facilitated by the \textit{transitional affordance} of the robotic benches, as they shifted seamlessly between being perceived as attention-grabbing robots that activated engagement from passersby, to exploration guides that distributed their engagement toward features and cues of the surrounding environment, and eventually settling into a mundane part of public infrastructure as their engagement naturally flowed into everyday activities. 

The contribution of our study is threefold. First, we presented the first documented design exploration, implementation, and in-the-wild deployment of a set of robotic benches in a public context. Second, we provided empirical evidence on how robotic furniture impacts the engagement of passersby within public space as an interactive urban intervention. Lastly, we proposed \textit{Affordance Transition Model (ATM)} that described the transitional affordances unique to public robotic furniture. Our study thus bridges two highly relevant yet still largely disconnected domains, robotic furniture and urban HCI, advancing ways to purposefully activate human experience with the built environment.

\section{Related Work}

\subsection{Public interactive interventions}
\hl{Urban design theorist \citet{carr1992public} has} pointed out that well-designed public spaces should support three essential human needs: (1) \textit{passive engagement} with the environment, where people observe others and the surrounding activity; (2) \textit{active engagement} through interaction with people or stimulation from the space itself; and (3) the excitement of \textit{discovery}, where novel or unexpected experiences enrich ordinary everyday life. Building on these foundations, prior work on public displays has shown how interactive systems can be deliberately designed to stimulate these needs, conceptualised by \citet{Memarovic2012DiscoveryPublicSpace} as the PACD model.
Complementing this,~\citet{Muller2010DesignSpace} conceptualise interaction with public displays as unfolding across phases of \hl{attracting attention, engaging and sustaining interaction, and various follow-up actions after disengagement}, and highlight design strategies that support smooth transitions between these states.

Furthermore, \citet{Fischer2012UrbanHCI} pointed out that urban HCI should not only focus on the digital interface itself but also on the spatial setup of public interventions. They provided a terminology of different spatial configurations that collectively shape how a public intervention is experienced.
Lastly, public interaction is also inherently social. \citet{Wouters2016HoneyPot} describe the \textit{“honeypot effect”}, showing how the presence of initial users around an installation attracts further bystanders, creating cascading forms of collective engagement.  

Work on public interactive interventions has primarily focused on introducing visual stimulation into the space, most often through screen-based displays~\cite{Parker2018Still,Ackad2016Skeletons}, projections~\cite{Hespanhol2015IntuitiveStrategies}, or lights~\cite{Lloret2014Puzzle}. Some works explored non-digital display approaches, for instance, chalk spray stencils used to visualise urban data directly on pavements~\cite{Koeman2015UrbanVoting} or using artificially created traces of use to contextualise public interfaces for meaningful urban interaction design~\cite{Hirsch2025}. Other work in this domain has further begun to investigate increasingly physical forms of displays. ~\citet{Hoggenmueller2019taxonomy} proposed a taxonomy of public displays along two dimensions: (a) the degree of physical integration with the environment, and (b) the mobility of the display. This framework highlights the potential of novel technologies, such as drones and urban robots, to distribute digital content in highly physicalised and ubiquitous forms. 


\subsection{Robotic furniture}
Building on the framework of implicit interaction~\cite{ju2015}, design researchers have investigated how existing everyday infrastructure and objects can communicate through motion, for example, interpreted as gestural cues that condition its approachability~\cite{ju2009approachability}. Since then, researchers have extended this perspective to different forms of robotic furniture. Much of this work has examined robotic furniture using gestures to signal intent or immediate affordances. For example, \citet{Sirkin2015Ottoman} explored how a mechanical ottoman could use gestures, such as approaching, nudging, or wiggling, to invite people to rest their feet, and to signal withdrawal when disengaging. The forward–back motion of a robotic chair toward a table has been shown to effectively recruit passersby to sit down and participate in a public chess game~\cite{Agnihotri2019PersuasiveChairBots}. Yet in the context of negotiating passage around a standing table, the same motion was perceived as a demand for right of way, rather than an invitation~\cite{Knight2017ChairBotGesture}. These contrasts illustrate that the meaning of robotic furniture motions is contingent on the spatial and situational setting in which they unfold.

Beyond facilitating immediate interaction, robotic furniture has also been investigated as a means of actively shaping social interactions among people. For example, \textit{MovemenTable} explored how moving tabletops could connect, separate, or reposition themselves to influence collaboration, guiding whether people came together, dispersed, or shifted between shared and individual activities~\cite{Takashima2015MovemenTable}. In ~\cite{Guo2023TalkwithStrangers}, a pair of bar stools was animated to rotate two occupants to face one another, subtly promoting conversation between two strangers. Furthermore, robotic furniture has been used to reshape layouts and mediate socio-spatial relations. \citet{Nguyen2022Responsive} explored how the autonomous reconfiguration of a robotic wall could shape occupants’ perception of place, showing how adaptations that blocked or exposed entrances, modulated visibility, or grouped furniture influenced circulation, privacy, and sociability in a shared space. Complementing this perspective, \citet{Nguyen2024Adaptive} introduced a mobile robotic partition that autonomously repositioned itself within an open-plan office, enabling workers to prevent, anticipate, or mitigate disturbances while simultaneously signalling social cues to colleagues. Furthermore, the recent work of \citet{Nguyen2025ArchitectonicGestures} has shown that the reconfiguration of robotic furniture itself can be designed to be communicative, introducing the notion of \textit{architectonic gestures} as movements that leverage the spatial impact of furniture to convey intent.

\subsection{Encountering robots in public spaces}
In attuning to how technologies shape urban life, the increasing presence of robots in everyday public spaces has sparked growing interest in the CHI and HRI communities to examine how they are encountered in situ. Some studies have focused on direct interactions, identifying a range of human responses such as offering assistance~\cite{Dobrosovestnova2022Help,Xinyan2025Breakable}, negotiating conflicts~\cite{Babel2022Reactions,Xinyan2024Understanding}, disturbing~\cite{Weinberg2023Sharing}, or even engaging in abusive behaviour toward robots~\cite{Kidokoro2015Escaping}. Some studies unfold encounters not only as direct interactions, but also in relation to the social and spatial surroundings in which they unfold, aligning with research interests in non-dyadic HRI~\cite{Schneiders2022NonDyadic,Sebo2020Teams}. For example, the ethnographic study by \citet{Pelikan2024Encountering} shows how delivery robots become woven into the socially organised life of the street, not merely navigating obstacles but blending with human activities, temporary work zones, and the mundane routines of urban life. Bystanders performed subtle forms of \textit{accommodation work}, stepping aside, adjusting their pace, or momentarily yielding priority, which allowed the robots to pass and, in doing so, rendered them as ordinary members of the street. The field deployment of robotic trashcans in a busy city square~\cite{Brown2024Public} positioned the robots not as entities merely passing through, but as \textit{of the space}, with their familiar form and functional tie to the square made them part of the city’s background infrastructure. Within this framing, people’s reactions emerged naturally and were described by the authors as Spontaneous Simple Sequential Systematics (SSSS) of interaction (e.g., an “offer and release” sequence where trash was presented to and then disposed of in the robotic trashcans). Extending these perspectives, \citet{Pelikan2025MakeSense} propose a framework, informed by sociological studies of placemaking~\cite{pps2003makes,whyte1980social}, to guide the design of urban robots in public environments. They identify four characteristics of public places that are critical for robot design, including localism, environments, activities, and sociability. This framework highlights that urban robots are not only actors in public space but also active participants in the creation of place. 

Alongside encounters with robots in functionally oriented deployments, another line of research has purposefully introduced robots into public spaces to create playful encounters that spark curiosity and delight. For example, \textit{Woodie}, a chalk-drawing robot deployed in a city laneway, invited passersby to co-create on the pavement, fostering playful forms of placemaking~\cite{Hoggenmueller2020Woodie}. \textit{BubbleBot}~\cite{Lee2019BubbleBot} released soap bubbles in public spaces to capture joint attention and nudge serendipitous encounters among strangers. Unlike traditional public displays and urban media interventions that primarily add digital layers to the environment, both works underscore how the physical form and embodied actions of robots can themselves become mediums of public experience.

\subsection{Summary}
Work on public interactive interventions has pointed to the possibility of deeper integration with their spatial context, where both form and content are interwoven to shape public experience. The ability of robotic furniture to reshape people’s spatial experience, \hl{along} with recent studies of urban robots \hl{showing} how physical form and embodied action can become a medium of public experience, speaks directly to this possibility. Together, these strands highlight the opportunity of animating existing infrastructure, such as the public bench, as a new way of shaping people's interactions in public spaces.

\section{Design process}

In our work, we followed a research-through-design approach~\cite{Zimmerman2007RtD}, a reflective design practice that begins with open-ended inquiry rather than addressing a predefined problem. Through a series of iterative explorations, including field observations, a lab-based design workshop, and an in-situ trial with passersby, we gradually refined our understanding of the role robotic benches can play in public space. It surfaced the potential of robotic benches to mediate socio-spatial engagement, revealed \hl{that such mediation emerges through the interplay between the robotic benches’ gestures and the surrounding spatial context}, and yielded practical insights into how such gestural strategies can be designed.


\subsection{Pre-deployment observation}
Our design process began with two of the authors observing how people use public benches across several locations where they were based, including parks, city centres, and residential areas. Following \citet{Gehl2013PublicLife}'s methods for studying public life, we mapped the physical layout and contextual features of the sites, traced people’s movement trajectories, and recorded their activities and surrounding social dynamics through illustrations and textual notes, complemented by photographs (Fig.~\ref{setup}, left). At the same time, as co-present observers, sitting outside ourselves, we recorded reflexive accounts of our own bench-sitting. 

Over the course of one week, these observations drew our attention to several salient patterns: (1) Passive sociality: Similar to prior studies of public sitting~\cite{Afonso172017bench,whyte1980social}, we observed strangers sharing a bench for extended periods without interacting, while a situational trigger (e.g., a crying baby) could suddenly spark conversation between them. (2) Disconnection between benches and their surroundings: People’s sitting could be undermined when benches were misaligned with situational needs or points of interest. For example, we observed one person leaving an empty shaded bench to sit unusually close to a stranger in the sun on a cold winter day, while in another case, people abandoned their seats to stand and watch a parade unfolding behind them. (3) Bench morphology shaping activity: People’s activities were strongly influenced by the physical design of benches, such as shape, orientation or presence of armrests. We observed people adapting their posture to these constraints, for example, reorienting their bodies around an armrest to maintain eye contact during conversation (Fig.~\ref{setup}, left). Together, these observations provided us with a more situated understanding of our design context. They also foregrounded two aspects for our \hl{initial design} investigation: how robotic benches might mediate social interactions among people, and how they might engage passersby with their surrounding space.


\subsection{Study site}
We chose to carry out our study in the semi-outdoor arcade area of a university library, shown in Fig.~\ref{setup} (middle, A). The building is not only a study place for students, but also a well-known tourist attraction and cultural heritage site of the city, where visitors frequently come for self-organised or guided tours. 
The chosen location is a transitional arcade area that connects the library interior with a large public square in front of it. The main pathway through the arcade passes three archways that directly face the library doors and is used by the majority of visitors entering and exiting the site, whereas the two side entrances are rarely used (see Fig.~\ref{setup}, middle). The arcade area is visible from both the public square via a decorated iron fence, and from the library interior through a transparent glass wall. Leveraging visibility and accessibility, the arcade has been historically used by both the university and the city as a public exhibition place. At the time of the deployment, the arcade hosted a large artistic installation involving a three-meter sculpture standing next to the library entrance.

\subsection{Robotic bench implementation}
\begin{figure*}[h]
\begin{center}
\includegraphics[width=1\textwidth]{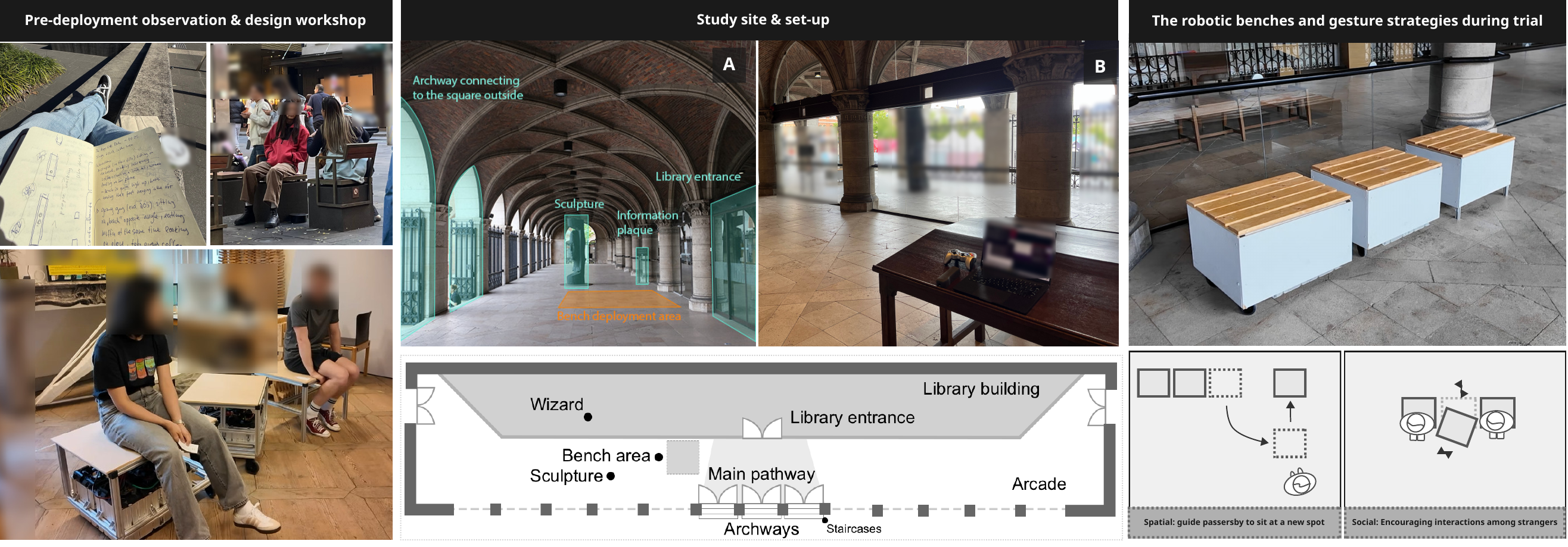}
\end{center}
\vspace{-8pt} %
\caption{Design process and study site. Left: observation notes and photographs from the design workshop. Middle: study site and setup with floor plan. Right: the robotic bench and gestural strategies probed during the trial.}\label{setup}
\Description{The figure shows the design process, study site, and robotic benches. On the left, photos depict pre-deployment activities, including field observation notes, sketches, and participants interacting with early bench prototypes during a design workshop. In the middle, images show the semi-outdoor library arcade where the benches were deployed, including the entrance, archways leading to the public square, a sculpture, and an information plaque, alongside a floor plan that marks the bench area, pathways, and wizard station. On the right, photos show three robotic benches arranged in the arcade, together with diagrams of gestural strategies: one illustrates benches repositioning to guide passersby to new seating spots (spatial engagement), and another shows benches reconfiguring to encourage social interaction among strangers (social engagement).}
\end{figure*}

We designed the benches (see Fig.~\ref{setup} left) to approximate the familiar appearance of public outdoor seating for passersby and decided to build three identical units so that they could reconfigure into different seating arrangements. Each unit adopted a cubic form (60 × 55 × 70 cm), with its four vertical faces fabricated from MDF panels and finished with a concrete-effect paint, and its top surface finished with a light wood panel to signal its affordance as seating furniture. These finishes were chosen to imitate the materiality of the existing urban furniture in the square surrounding the library \hl{to preserve infrastructural legibility, as our design inquiry did not seek to explore variations in form but instead focused on designing gestures.}


We deliberately chose to animate only two of the three bench units, leaving one anchored as a static reference, so reconfigurations were more predictable, a property shown to foster more trustworthy engagement between unfamiliar occupants and robotic furniture~\cite{Nguyen2022Responsive}. Similarly, we purposefully hide the mechanical parts of the robotic units to improve their perceived robustness and stability, motivating passersby to confidently sit on them like on conventional public furniture.

The mobility of each robotic bench was enabled by the KELO Robile robotic platform\footnote{KELO Robile: kelo-robotics.com}, which was operated through the Robot Operating System (ROS). Using customised Python software modules, motion sequences of individual bench units, initially controlled by a researcher via a gamepad, were recorded, stored, and then executed via a laptop throughout the deployment. The study followed a Wizard-of-Oz approach~\cite{dahlback1993wizard}, with the wizard operating the bench units remotely from inside the library, positioned behind a glass wall overlooking the arcade area (see Fig.~\ref{setup} left, B). This setup allowed the wizard to maintain an unobstructed view of the deployment site while remaining invisible to passersby, as the reflections on the glass wall concealed their presence. 

\subsection{Design workshop and in-situ trial}

To explore the appropriate gestural strategies that the benches could employ, we conducted a design workshop with five participants, consisting of three authors and two researchers from the broader research team (all with a design background). The workshop took place in an open space at our lab, where we role-played some scenarios identified from the observations, such as strangers sharing a bench. In these scenarios, participants alternated between enacting passersby in public space and controlling the robotic benches via a gamepad to experiment with how the benches might intervene in public experiences (e.g., breaking the ice between strangers, or orienting attention toward a particular direction).   

Given the strong spatial and contextual dependencies of our investigation, we then moved from lab-based exploration of bench gesture to in-situ design exploration at the actual study site. Here, we trialled the previously developed gestural strategies with passersby, probing how robotic furniture could actively mediate both spatial engagement and social interaction in public seating. First, we wanted to explore ways for the robotic bench unit to approach and make passersby notice it. To do so, we controlled the robot to attract passersby using two gesture strategies~\cite{Nguyen2025ArchitectonicGestures}: (1) positioning one bench unit along the pathway, accompanied by an emblematic gesture resembling a playful “wiggle”; and (2) a deictic gesture in which the bench approached on a linear path from a distance. This also became our recruitment strategy. Once passersby noticed the bench and displayed interest, a researcher approached them and invited them to experience two additional gestural sequences designed to mediate social and spatial engagement.

Participants then experienced the two additional gestural sequences, which were designed as follows: (1) one bench unit guided people to a seating location by approaching them, rocking forward and backward twice, and then moving away to settle in a new spot; and (2) the middle bench unit broke the ice between strangers by moving forward and wiggling to prompt interaction (Fig.~\ref{setup}, left). One of the researchers enacted the role of a stranger seated on the bench, introduced to study participants as another participant, but did not initiate any interaction unless the participant engaged with them during the study. To scaffold their engagement, we provided short scenario prompts, such as \textit{“You are visiting the library and looking for a place to sit down and rest”}, and asked them to respond to the robotic bench naturally. Afterwards, participants were invited to a short \hl{semi-structure} interview \hl{on their interpretation and perception of the gestural sequences and their experience in such situations}. We video-recorded their interactions and audio-recorded the interviews. In total, the four-day trial involved 35 participants (P1–P35).

\subsection{Lesson learned}
First, the on-site trial allowed us to gain a deeper understanding of the study site. As the deployment took place during the university holiday period, tourists constituted the majority of library visitors. The arcade area therefore primarily functioned as a transitional space, carrying the majority of visitors in and out of the library rather than serving as a place to linger. 

In terms of attracting passersby’s attention, the stationary wiggle gesture was often overlooked, even when positioned near attentional bottlenecks (i.e., the main pathway to the library entrance). In contrast, the approaching robotic bench unit was more effective in recruiting passersby, as it drew greater attention and encouraged them to pause. This could be because the deictic gesture, with its larger physical amplitude, was more easily registered in peripheral vision than the subtle stationary wiggle. At the same time, its longer duration extended the temporal window for noticing, making it more likely to capture the attention of passersby whose focus in public space is often fragmented.

For robotic furniture mediating social interactions, although social exchanges emerged in most cases (n=31), interview responses revealed mixed perspectives. Seven participants suggested that such interventions seemed unnecessary or risked making the social dynamic feel \textit{“unnatural”}~(p6). Three participants further indicated that people’s social needs in public space are dynamic. While the robotic furniture's mediation could be helpful when they were open to socialising, such direct interventions were perceived as bothersome when they preferred to be alone. One participant expressed a particularly strong concern, asking, \textit{“Why does the bench want to tell me what to do?”} (P21), questioning the appropriateness of robotic furniture nudging human social interactions in a direct way.

For robotic furniture mediating spatial engagement, participants generally recognised the gesture as an invitation to follow the bench and sit in a new spot (n=29). Interviews revealed that such spatial mediation could have felt more meaningful when the new position was legible in relation to the surroundings. While nine participants noted ambiguity about why a particular location was chosen, others actively drew spatial inferences to interpret the behaviour of the bench. Three participants associated the new spot with the archway, as it faced the archway and offered a clearer view of the outside square, and they interpreted the behaviour as an invitation to sit there for a \textit{“better look at the outside”} (P8). P11 further elaborated, describing the projected intention as fostering \textit{“a connection with the outside.”} This highlights a need for gestures to more explicitly cue their relation to the broader spatial context. 
Furthermore, these insights point to a bidirectional activation between the robotic benches and the spatial context. While the gestures shaped how people engaged with the surrounding space, those spatial elements, in turn, activated and re-framed people’s engagement with the benches.

\section{In-the-wild study}
The design process not only provided insights into gestural strategies for robotic public furniture, but also \hl{shifted our focus from mediating social encounters to their broader socio-spatial implications.} Building on these insights, we next conducted an in-the-wild study to further investigate how robotic benches and their surroundings mutually shape social and spatial engagement in public space.



\subsection{Socio-spatial activation gestural sequence}

\begin{figure*}[h]
\begin{center}
\includegraphics[width=0.9\textwidth]{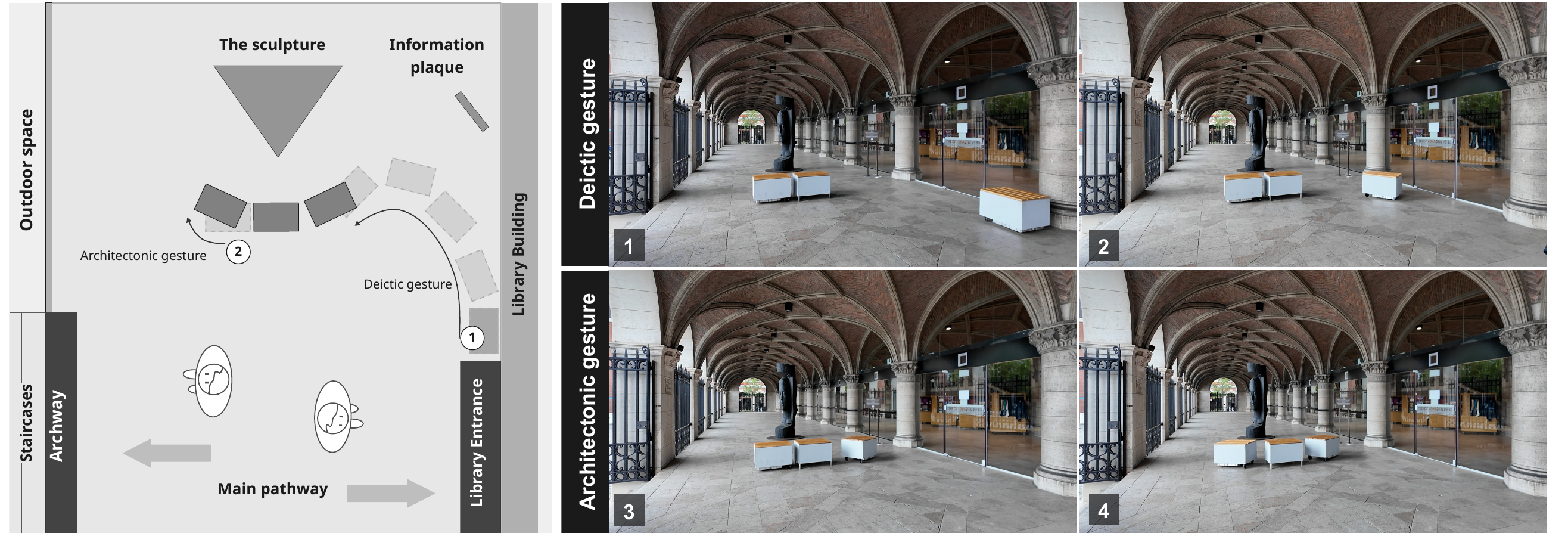}
\end{center}
\vspace{-8pt} %
\caption{Socio-spatial activation gestural sequence. Left: bird’s-eye view of the gestural trajectory in relation to the sculpture and study site. Right: screenshots of the robotic benches’ gestural sequence, consisting of two parts: (1) a deictic gesture moving toward the two other benches, and (2) an architectonic gesture that subtly indexed the sculpture as a focal point.}\label{concept}
\Description{The figure illustrates the socio-spatial activation gestural sequence of the robotic benches. On the left, a bird’s-eye floor plan shows the benches’ movement trajectory in relation to the library entrance, archways, main pathway, sculpture, and information plaque. Two types of gestures are annotated: a deictic gesture that directs attention toward the other benches, and an architectonic gesture that positions the sculpture as a focal point. On the right, a sequence of four photographs from the arcade captures the benches performing these gestures. Frames 1–2 show a bench moving toward the others (deictic gesture), and Frames 3–4 show the benches arranged symmetrically in front of the sculpture (architectonic gesture), subtly anchoring it as a point of focus within the space.}
\end{figure*}

We decided to leverage the existing sculpture, which had previously remained peripheral to most passersby, and designed the gestural sequence in relation to it to purposefully evaluate the bidirectional activation between the robotic benches and the spatial context. Two researchers further prototyped and experimented with gestures from the previous trial, ultimately deciding on a gestural sequence that combined two strategies~\cite{Nguyen2025ArchitectonicGestures}, including a deictic gesture to attract attention while also conveying directionality, and an architectonic gesture that leverages spatial impact to nudge engagement. 
Specifically, we first positioned one robotic bench unit near the library entrance, where passersby's attention was naturally concentrated. When an individual or a group of passersby arrived along the pathway, this unit would start performing a \textit{deictic} gesture by moving toward the two other benches previously placed in front of the sculpture, eventually forming a row of three. Once the three units were aligned, the two end units would start rotating approximately 20° inward, creating a gentle arc that oriented the cluster toward the centre (see Fig.~\ref{concept}). In doing so, the benches subtly indexed the sculpture as a focal point via an architectonic gesture. At the same time, the inward rotation followed the principles of f-formations~\cite{Kendon2010SpacingOrientation}, establishing a shared o-space that afforded joint attention and invited group sociability. Beyond attracting passersby engagement with the benches, this gestural sequence thus aimed to activate both spatial engagement with the sculpture, as well as social interaction between multiple passersby.
To ensure safety while also allowing sufficient time for the movements to capture bystanders’ attention, we set the benches’ speed to a constant 18 cm/s.

\subsection{Data collection}
The field study followed a Wizard-of-Oz approach~\cite{dahlback1993wizard}, with a setup similar to that of the \hl{in-situ} trial (see Fig.~\ref{setup}, left B). Two researchers were stationed inside the university library, behind a glass wall that overlooked the arcade area. The reflective surface of the glass concealed their presence from people outside. Once visitors appeared at the arcade area or were about to exit the library, the wizard remotely triggered the pre-recorded motion sequences from the laptop. The wizard purposefully allowed the interaction between the passersby and the benches to unfold naturally, until they either showed signs of leaving the area or, after sitting for a while, shifted their focus of attention away from the benches and back to their own activities. At that point, one researcher approached and conducted a short semi-structured interview with those who had interacted with, or noticed, the robotic bench. After each instance, the benches were repositioned to their initial configuration. To reduce biases from passersby who might infer the study procedure by observing the previous gestural sequence or the following interview, we deliberately left a temporal gap \hl{of approximately 5 minutes} before triggering the next sequence.

A camera was set up approximately 5 meters from the bench configuration location to video record individuals passing by or interacting with the robot
. Over a total of 26 hours of deployment across five days, the wizard triggered the benches 98 times: 14 (14.3\%) instances went entirely unnoticed, 51 (52.0\%) drew attention without leading to sitting, and 33 (33.7\%) resulted in passersby sitting on the bench. From these instances, we conducted 55 semi-structured interviews with 1 individual and 54 groups\footnote{This distribution is due to the site dynamics, as visitors typically arrived in groups and were more likely to participate interviews.} of people who interacted together (F1-F55). \hl{There were a total of 148 people in these interview groups, with group sizes ranging from 1–6 $(M=2.82,\ SD=1.12)$}. \hl{The rest either left too quickly to be invited or declined the interview.} The interview questions focused on unpacking the participants’ engagement with the benches and surrounding space, their interpretations of the gestural sequence, and their overall experience. 
As passersby in public spaces are often in a hurry, interviews were kept short and concise, typically lasting 2–5 minutes, similar to comparable studies on in-the-wild interactions with robots~\cite{Xinyan2025Breakable, Hoggenmueller2020Woodie}.

The study followed a protocol approved by the university’s Human Research Ethics Committee and complied with local regulations on data collection and the incidental recording of individuals in public spaces. Participants were approached after they had either interacted with, or noticed, the robotic bench, and their consent was obtained prior to participation in interviews. In line with similar practices for field research in public spaces~\cite{Brown2024Public}, we did not seek consent from passersby who were incidentally included in the video recordings; however, measures were taken to protect their privacy by blurring identifiable features.

\subsection{Data analysis}
We employed a cross-analysis approach, where interview data offered deeper insights into and explanations of passersby behaviours observed during the field study; video recordings enriched the contextual understanding and triangulated participant comments. \hl{The first author led the analysis, drawing on their background in design research and human–robot interaction in public space, which sensitised the analysis toward design-led interpretations of how people made sense of and engaged with robotic artefacts. The co-authors' expertise in architecture and robotic furniture design provided complementary readings that foregrounded spatial configurations and environmental relations.} 

Following a bottom-up approach, the first author closely examined the video recordings and documented passersby behaviours, which were captured through textual descriptions and supplemented with screenshots. Subsequently, an approximately one-hour meeting was held between all authors, to discuss and review the emergent interaction patterns identified by the first author. The interview data were transcribed and analysed using thematic analysis~\cite{braun2006thematic, Braun2019reflexive}. The first author coded the data inductively, identifying recurring patterns from which sub-themes were derived. This initial coding scheme was then discussed and refined during a one-hour meeting with all authors. Subsequently, the sub-themes were deductively organised into final themes, structured around two main aspects of our investigation: (1) supplementing our interpretations of how passersby engaged with the robotic benches and the surrounding space, and (2) understanding how passersby interpreted the intentions of the robotic benches. \hl{During this iterative process, the coding scheme\footnote{The full initial and final coding schemes are provided in the supplementary material.} evolved through abstraction and consolidation. For example, a set of behavioural codes were later grouped under the overarching \emph{Sitting} category, which, drawing on an analytical lens informed by the PACD model, was refined into \emph{discovery}, \emph{active}, and \emph{passive} forms of sitting.} \hl{Our} approach aligns with reflexive thematic analysis~\cite{Braun2019reflexive}, \hl{treating coding as an iterative and interpretive process rather than a reliability-focused categorisation. Counts reported in the Results section are provided to contextualise the prevalence of observations and do not imply frequency-based validation. }



\section{Results}

In this section, we first draw on observational data to examine how the robotic benches shaped interactions in situ. We begin with direct engagements between passersby with the benches, then broaden the lens to consider how the benches activated their engagements with the surrounding space, and finally analyse the social interaction they nudged between passersby. Interview insights directly related to these observations are incorporated to aid interpretation. Lastly, we report insights from interviews that reveal the interpretive frames through which participants perceived the intentions of the benches.

\subsection{Engagement with the robotic benches}
Our participants appropriated sitting as a way of engaging with the robotic benches, showing different levels of discovery, active, and passive engagement depending on where and when they sat. During discovery engagement, they also employed other embodied gestures to assert control over the benches.

\subsubsection{Sitting as discovery, active, and passive engagement}
\label{sec:sittingPACD}
\begin{figure*}[h]
\begin{center}
\includegraphics[width=1\textwidth]{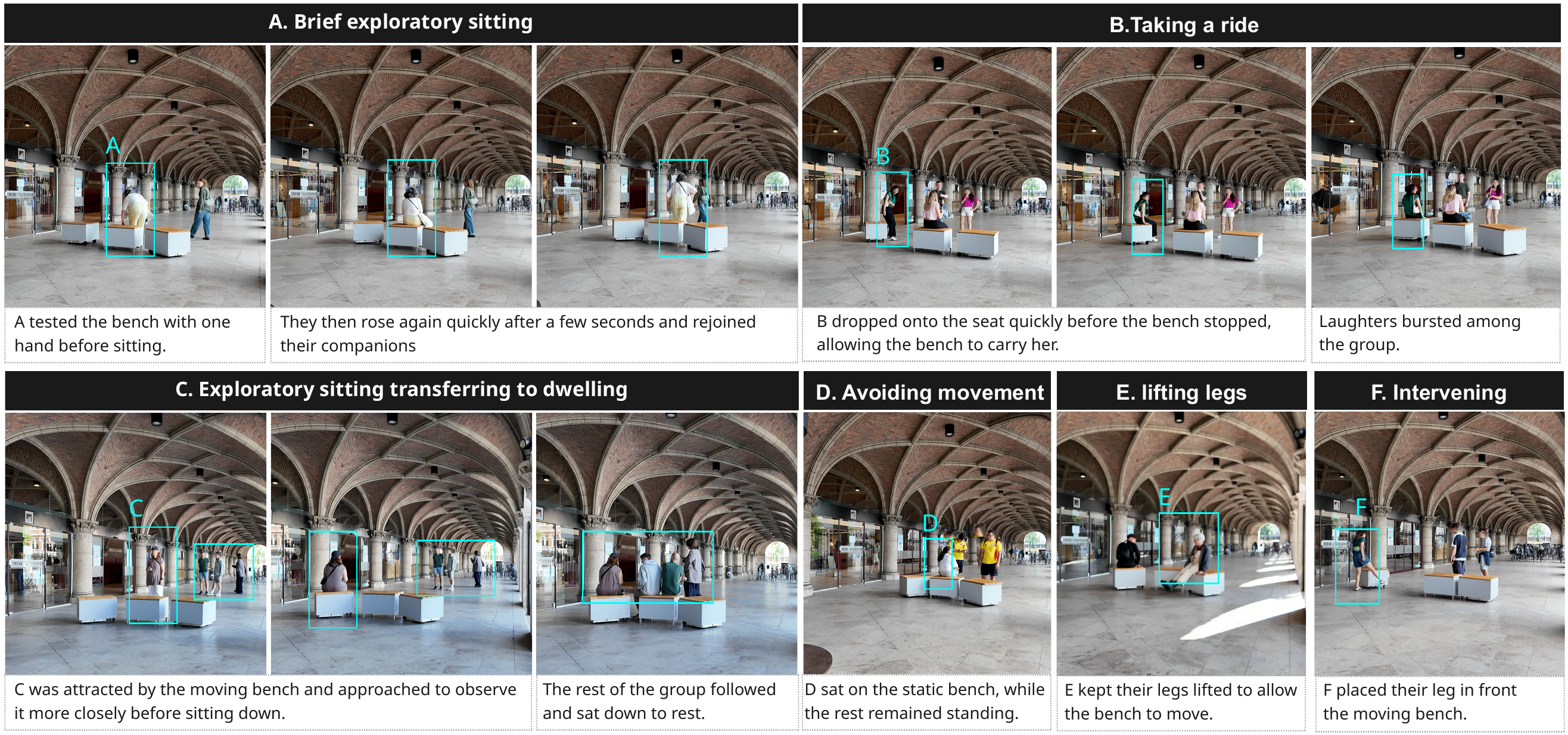}
\end{center}
\vspace{-8pt} %
\caption{Instances of interactions with the robotic benches. A: Brief exploratory sitting; B: Taking a ride; C: Exploratory sitting transferring to dwelling; D: Avoiding movement; E: Lifting legs; F: Intervening}\label{withbench}
\Description{The figure presents six categories of how passersby interacted with the robotic benches, illustrated through photographic sequences. A. Brief exploratory sitting: A participant briefly tested the bench by touching it with one hand and sitting for only a few seconds before rejoining companions. B. Taking a ride: A participant sat down just before the bench stopped, allowing it to carry her a short distance, provoking laughter among her group. C. Exploratory sitting transferring to dwelling: One participant, attracted by the moving bench, approached and sat down, followed by the rest of the group, extending their stay. D. Avoiding movement: A participant chose a static bench, remaining seated while others in the group stayed standing. E. Lifting legs: A participant lifted their legs to allow the moving bench to pass beneath without obstruction. F. Intervening: A participant placed their leg in front of the moving bench to deliberately interrupt its motion.}
\end{figure*}

Most passersby (\(n=22, 22.4\%\)) who chose to sit down on the robotic bench after its gestural sequence were motivated by the sense of curiosity rather than the need to rest, resulting in a brief period of sitting for only a few seconds as a form of \textit{discovery} engagement.
Because of its dynamic motion, passersby typically waited until the bench came to a halt before sitting down. Several displayed visible caution, reaching out a hand to test its stability before committing to sit. Fig.~\ref{withbench}, A, illustrates such an encounter: a passerby tentatively extends a hand toward the bench’s surface, testing its steadiness before cautiously lowering themselves onto it. Moments later, they rise again and rejoin their companions without lingering. 
During the interview, participants confirmed being intrigued by the unusual sight of a bench that moves, describing that "[this] made them want to \emph{“try it out.”} (F38) Twelve participants further explained that they deliberately sat down to see \textit{“if it was gonna move when [they] sit down.”} (F9).

In four cases (two child cases and two adult cases, \(12\%\)), participants chose to sit down on the robotic bench during its gestural sequence performance as a particularly playful \textit{discovery} engagement. As shown in Fig.~\ref{withbench}, B, passerby C quickly dropped onto the seat, allowing the bench to carry her, moving closer to a friend already seated on another bench. The ride prompted bursts of laughter among the group. F32, who also rode the bench, highlighted the playful nature of the experience, describing themselves as feeling \emph{“like a cat sitting on top of a robotic vacuum”} when seated on the moving bench.

In five cases (\(5.1\%\)), discovery sitting transformed into a more sustained \textit{active} engagement with the benches as a place to rest, once participants perceived the robotic benches, after remaining still for a period, as offering the identical affordances as the non-robotic one. For instance, in Fig.~\ref{withbench}, C, a group of passersby on their way to the library paused when one member, intrigued by the gesturing bench, diverted from their original path to observe the situation more closely. They then sat down and called the rest of the group, who also joined in sitting. As they settled, their attention gradually shifted away from the novelty of the robotic bench toward more everyday activities: resting, chatting, and checking their phones. 

In contrast, seven passersby (7.1\%) chose to only \textit{passively} engage with the robotic benches, as they intentionally sat on the middle bench after recognising that it did not have wheels and would remain stationary. As one participant explained, they \emph{“chose the safe option”} (F12). An illustrative instance of such more cautious behaviour can be seen in Fig.~\ref{withbench}, D, where the passersby opted for the static unit (later confirming in the interview that this choice was deliberate), while their companions chose to remain standing and chat rather than sitting on the two mobile benches.

\subsubsection{Gesturing as discovery engagement}
Because the robotic bench did not offer an explicit interface for interaction, 11 participants (\(11.2\%\)) improvised embodied gestures as a form of discovery engagement, attempting to assert control over its robotic behaviour.
After waiting for the bench to stop and sitting down, seven participants deliberately lifted their legs off the ground to enable the bench’s mobility (See Fig.~\ref{withbench}, E), an action one participant described as gesturing the bench to \emph{“take us for a ride”} (F49). In contrast, four participants physically intervened in the motion of the bench by placing their leg in front of it (see Fig.~\ref{withbench}, F). In three of these instances, participants positioned their leg firmly against the front side of the bench as it advanced, maintaining contact even as the bench continued to push forward, until it eventually came to a halt. During the interview, two participants explained that they were testing whether the bench would sense their presence and stop automatically. The other two cases instead occurred on a particularly windy day, when participants misinterpreted the gestural sequence as motion caused by the wind and intervened by blocking the bench with their leg to “save” the situation.


\subsection{Engagement with the surrounding space}
The gestural sequence of the robotic benches activated the engagement of passersby with the surrounding space at varying levels, from passive to active and discovery. In turn, the spatial context shaped the way they engaged with the robotic benches.

\subsubsection{Robotic benches activating spatial engagement}
\label{sec:spillover}
\begin{figure*}[h]
\begin{center}
\includegraphics[width=1\textwidth]{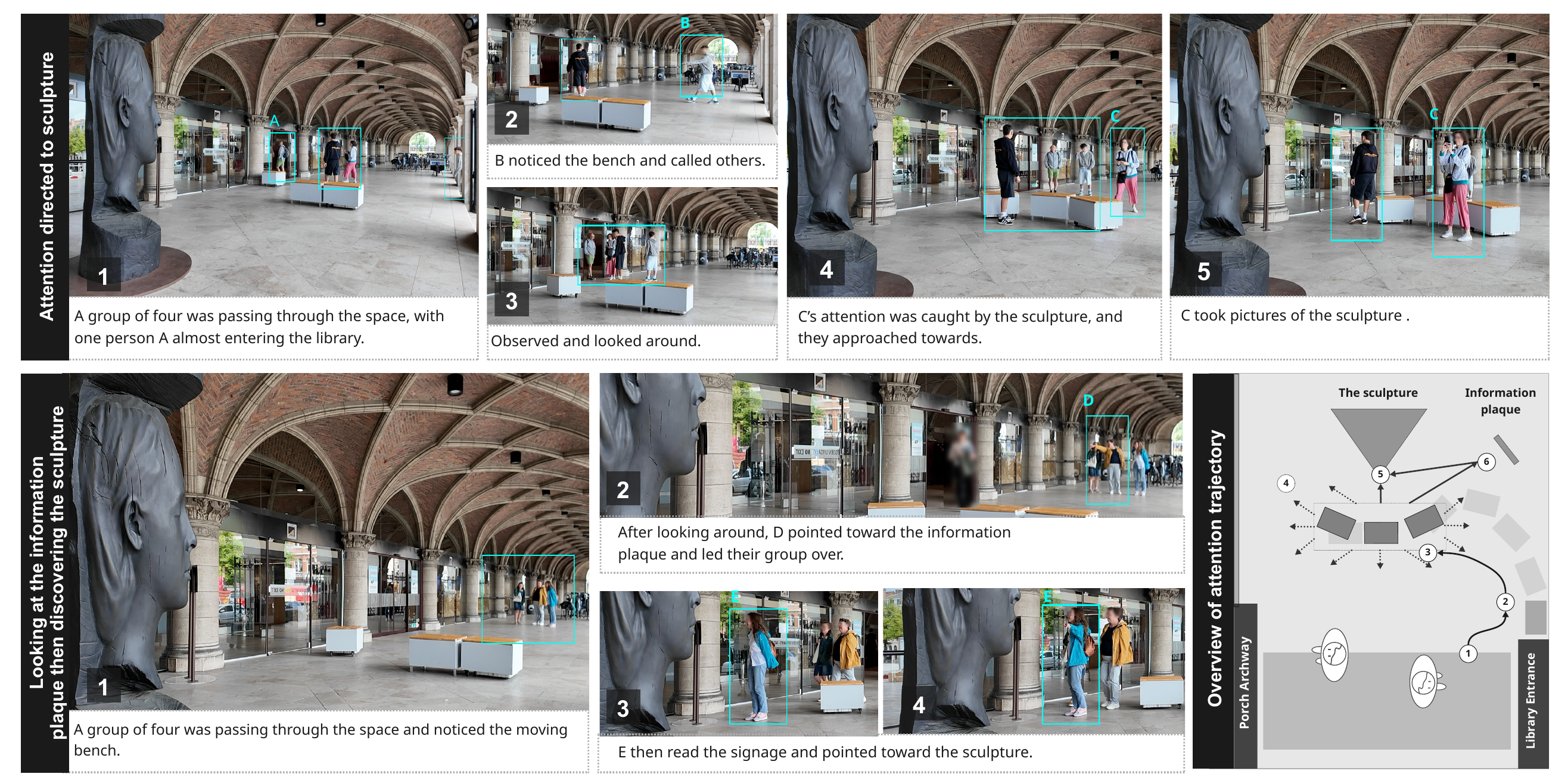}
\end{center}
\vspace{-8pt} %
\caption{Passersby’s serendipitous encounters with the sculpture prompted by the robotic bench. Top: a group noticed the bench and shifted attention to the sculpture. Bottom left: another group first examined the information plaque before turning to the sculpture. Bottom right: overview of passersby attention trajectories, beginning with passersby entering or exiting the space (1), noticing the moving bench (2) with later joined the other units (3), spilling attention over to scan the surroundings for cues (4), and then their attention either landed at the sculpture (5) or first to the information plaque (6) before bouncing back to the sculpture.}\label{Attention}
\Description{The figure illustrates how passersby’s attention was redirected from the robotic benches toward the nearby sculpture, captured in two narrative sequences and one trajectory diagram.
Top row (Attention directed to sculpture): A group of four initially walked through the arcade, one almost entering the library. After one participant noticed the robotic bench and called others, the group paused, looked around, and eventually had their attention caught by the sculpture, leading them to approach it and take pictures.
Bottom left (Looking at the information plaque then discovering the sculpture): Another group first noticed the moving bench, then one participant pointed toward the nearby information plaque. After reading the signage, another participant gestured toward the sculpture, shifting the group’s focus to it.
Bottom right (Overview of attention trajectory): A diagram maps attention flow: passersby entered, noticed the moving bench, spilled attention across the space in search of cues, and then redirected their focus either toward the sculpture or via the information plaque before returning to the sculpture.}
\end{figure*}

While most library visitors typically passed through the arcade area without lingering, the robotic benches effectively motivated them to pause, divert, or return, prolonging their \textit{passive} engagement with the space. In 83 cases (\(84.7\%\)), when passersby noticed the benches in motion, they often paused to observe them. Over half of these passersby (\(n=51, 52.0\%\)) deliberately diverted from their original trajectory to approach the benches more closely, shifting from the main pathway toward the side of the arcade. In four cases, individuals who had already left the arcade area, either by entering the library or exiting through the archway, even returned to the space specifically to observe the benches. As shown in the top row of Fig.~\ref{Attention}, A, who had already stepped into the library entrance, turned back into the arcade area toward the benches for a closer observation after their companions noticed the bench movement. These moments of pausing, diverting, and returning thus extended the passive engagement of passersby with this space from what was otherwise a brief encounter.



During discovery engagement with the robotic benches, most passersby deliberately scanned the surroundings to search for cues to to explain or contextualise the gestural sequence, resulting in a spillover of attention that sometimes fell on the nearby sculpture (\(n=13\)) or its information plaque (\(n=7\)), activating their \textit{active} and \textit{discovery} engagement with it. During interviews, participants confirmed that they had not originally planned to look at the sculpture, but were drawn to it after noticing the bench. For instance, in Fig.~\ref{Attention}, top row, the group looked around after spotting the moving benches, seemingly seeking further clues from the surroundings. This led C (F14) to focus on the sculpture and eventually walk toward it for closer appreciation and photographs.
Other participants, who assumed that the information plaque of the sculpture was to explain the gestures of the benches, inadvertently discovered the sculpture through it. One illustrative example is shown in Fig.~\ref{Attention}, bottom row. After noticing the bench, D pointed toward the information plaque and guided their group over. E then read the signage and pointed towards the sculpture, which eventually drew the group to shift their attention to and examine the artwork.



Across these encounters, the robotic bench operated not only as an actor of direct interactions, but also as an intermediary, mediating attention flow, redirecting passersby to engage with the cultural artefact that might otherwise have been overlooked. Importantly, this redirection was activated in a subtle and almost unnoticeable way. As F12 reflected, this subtle activation approach was actually more effective than more explicit guiding gestures, because \textit{“as humans, we don’t follow stuff that is too obvious.”}

\subsubsection{Spatial context shaping robotic engagement} \label{sec:orientation}
In two cases (\(2\%\)), passersby who entered the arcade area with the initial explicit intention of viewing the artwork chose to engage actively with the robotic benches, as they thought that the benches were part of the artistic installation.
For example, a couple (F37) approached the sculpture directly without initially noticing the bench. After the bench began to move, they turned their attention toward it and interpreted its trajectory of moving closer to them as \emph{“reacting to us”}. Feeling invited to sit, the couple prolonged their engagement with the sculpture. Later, in the interview, they reflected that without the bench’s movement, they would not have chosen to sit down, and their encounter with the sculpture would have been much shorter.


Whereas nine participants (\(9.2\%\)) chose to sit on the benches facing the sculpture as the design intended, 24 other participants (\(24.5\%\)) chose to face another direction because of their own personal space-use preferences.
In seven instances, even after noticing the sculpture and recognising that the seating arrangement was oriented toward it, participants nevertheless decided to face the pathway rather than the sculpture. F28 explained that they were ~\emph{“waiting for friends,”} which made it practical to face the pathway. During the study period, an event was also taking place in the square just outside the arcade, creating another focal point of attention. F44, for instance, described their decision not to face the sculpture by noting a preference to watch a nearby crane dismantling event structures: \emph{“We were watching the crane. That’s impressive too.”}

\subsection{Engagement in a social context}\label{sec:social}
The gestural sequences of the robotic benches nudged shared amusement within groups and encouraged social connections between strangers, yet at the same time, created unintended social pressure for individuals seeking solitude.

\subsubsection{Nudging social amusement}
Upon noticing the gestural sequence of the robotic benches, a passerby who arrived in a family or friend group often had the urge to share their amusement from the benches with their peers, who in turn empowered them to actively engage with the benches. This was exemplified by members pausing and turning to their peers to comment on the benches, sometimes bursting out with exclamations of surprise or collective laughter. In cases where not everyone had noticed the gestures, those who did often drew the attention of their companions to join in the engagement. For example, as shown in Fig.~\ref{Attention}, B, who was the only one to notice the bench movement, pointed towards the bench and called over their family members to observe the bench together. 

\subsubsection{Nudging social connection}
\begin{figure*}[h]
\begin{center}
\includegraphics[width=1\textwidth]{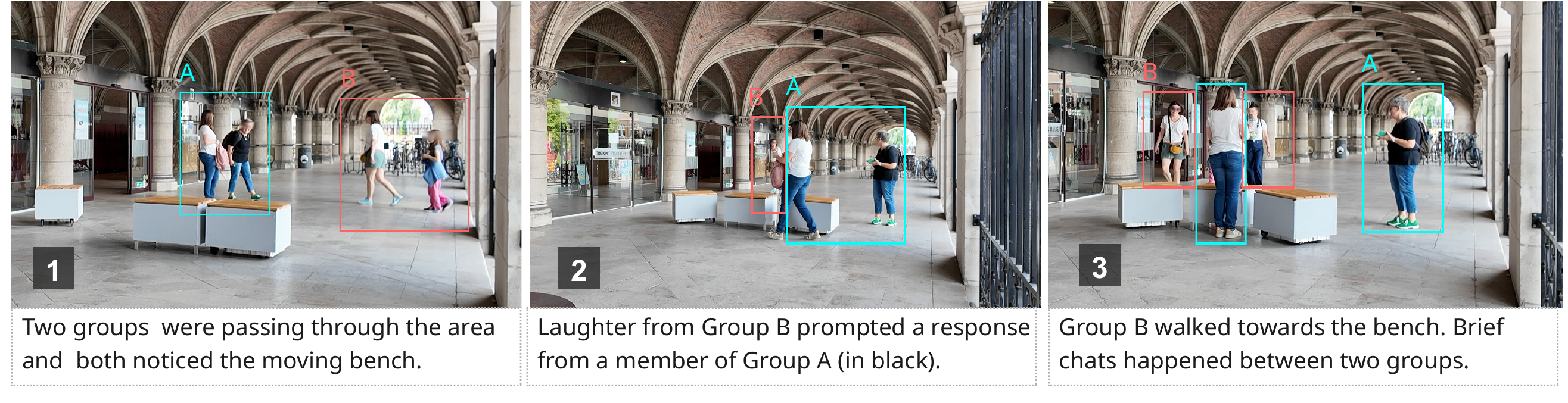}
\end{center}
\vspace{-8pt} %
\caption{Sequence of interaction showing how the bench prompted brief social exchanges between two groups.}\label{social}
\Description{The figure shows two groups of passersby whose attention to the robotic bench led to spontaneous interaction.
Image 1: Two groups walking through the arcade simultaneously notice the moving bench.
Image 2: Laughter from Group B elicits a playful response from a member of Group A, dressed in black.
Image 3: Group B approaches the bench, sparking brief chats between members of the two groups.
The sequence illustrates how the robotic bench catalysed unscripted social interaction between previously unconnected groups.}
\end{figure*}

The urge to share amusement and surprise at the robotic benches sometimes extended beyond immediate groups to reach across multiple groups that are unfamiliar with each other, prompting group members to initiate brief social exchanges with strangers (\(n=5)\) while engaging together with the robotic benches.
These social exchanges ranged from subtle social gestures, such as brief eye contact or a smile towards each other to acknowledge the shared noticing of the bench, to short verbal exchanges where people commented on what they saw or sought clarification from each other about what was happening. For instance, in Fig.~\ref{social}, two groups (A and B) were passing through the area at the same time when both noticed the moving bench. The laughter from Group B prompted a response from a member of Group A (the woman in a black skirt), who smiled back towards Group B in acknowledgement (see Fig.~\ref{social}, 2). Then, Group B walked towards the bench for closer observation. As both groups gathered around the benches, this brief encounter unfolded into short chats accompanied by shared laughter. Such instances illustrate how the robotic bench functioned as a social catalyst that momentarily connected strangers in public space.

\subsubsection{Nudging unintended social pressure}
\label{sec:spotlight}
\begin{figure*}[h]
\begin{center}
\includegraphics[width=1\textwidth]{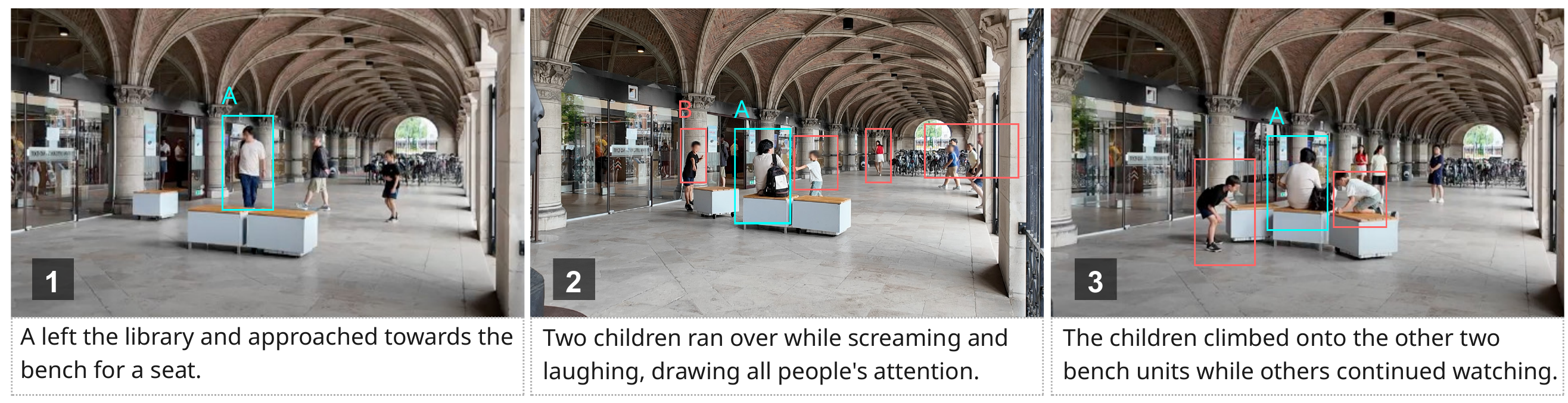}
\end{center}
\vspace{-8pt} %
\caption{Sequence of interaction showing how a person seated on the bench became the unintended centre of attention for surrounding passersby.}\label{spotlight}
\Description{Accessibility Description:
The figure shows how the robotic benches drew playful attention and engagement from both adults and children. Image 1: A person exits the library and approaches the robotic bench to sit. Image 2: Two children run toward the benches, screaming and laughing, which captures everyone’s attention. Image 3: The children climb onto the remaining bench units, while others nearby watch the unfolding activity.This sequence highlights how robotic benches not only attracted individual use but also created moments of collective attention and playful engagement.}
\end{figure*}

For individuals who sought solitude in public spaces, the social affordances of the robotic benches in nudging amusement and connection inadvertently created social pressure to them, turning their passive engagement into a negative experience.
In one instance, F55, who had just left the library, sat down on the middle non-robotic bench to rest (Fig.~\ref{spotlight}, A). While he was seated, the other two benches were still in motion to complete the gestural sequence, attracting the attention of many passersby. This inadvertently placed F55 in the spotlight, as others observed the benches approaching him, as shown in Fig.\ref{spotlight}, (2), 
placing him at the unintended centre of focus. The situation was further intensified when two children, drawn by the moving benches, ran over while screaming and laughing. They climbed onto the other two bench units, which amplified the attention. Later, in the interview, F55 expressed discomfort about unintentionally attracting attention, describing the experience as \emph{“bothering”} when he had simply wanted to rest quietly and mind his own business, and adding that he felt \emph{“a little bit afraid it will attract too many strangers to this side.”}





\subsection{Interpreting the intention of the robotic benches}
\label{sec:Interpret}
Our participants interpreted the autonomous intention of the robotic benches based on multiple interrelated factors, including their physical design, collective architectonic arrangement, situatedness within the surrounding socio-spatial context, gestural timing, and resemblance to other familiar robots.

\subsubsection{Physical morphology}
\label{sec:physical}
Participants interpreted the familiar, mundane physical design that blended into the surroundings of the robots as an invitation to sit on, with 41 passersby noting that they recognised them as benches intended for sitting.

As F2 noted: \textit{“it just looks like a bench, isn’t it”}, F13 further pointed out that its ordinary look, rather than appearing as \textit{“something super cool”}, allowed the benches to \textit{“fit into the background,”} which in turn enabled them to connect the benches to other elements of the space and notice the nearby sculpture. Nine participants made closer observations of the benches’ material details and noted that the middle bench lacked wheels, which shaped how they engaged with the bench (e.g., deliberately choosing to sit on the non-moving unit).

\subsubsection{Architectonic arrangement}
Participants interpreted the symmetrical, balanced arrangement in the final architectonic gesture of the three bench units as conveying not only their belonging to a shared group, but also their intention to create a social space for multiple people to sit. 
Eleven participants interpreted the deictic gesture of the robotic bench moving from the library entrance as “return[ing] to its original position” (F18), or, as F11 described, \textit{“to meet the other bench”}.
In contrast, five participants initially perceived this gesture as ambiguous, which was then clarified once the benches aligned symmetrically, which conveyed a stronger sense of spatial intentionality: \textit{“we didn’t understand what it would do, then we saw the second one also moving, so it was for us”} (F8). 
Eleven participants further interpreted the architectonic arrangement as socially oriented, with the subtle arc created by the inward rotation of the benches seen as deliberately shaping a shared space for a group.   

\subsubsection{Socio-spatial context}
\label{sec:interpretatingcontext}
Situated within the socio-spatial context of a university library, the robotic benches were interpreted by participants as intending to perform different public functions.
Some assumed that the benches might be used for delivering books for the library (n=5). Others drew on the cultural atmosphere of the library as a place that hosts artworks, interpreting the benches as an independent art installation (n=12). As F11 remarked, \textit{“we thought it was a piece of art, because there are several artworks here.} Eleven participants explicitly linked the benches to the nearby sculpture, interpreting them as a deliberate attempt to direct attention toward it. As F10 remarked, it was \textit{“kind of a nudge towards the sculpture.”} Others interpreted the bench behaviour in relation to the wider layout of the arcade, for example, suggesting that the benches were deliberately lining up to form a \textit{“barrier”} (F48) to prevent people from moving further down the side of the space.
Two participants, however, expressed concerns about safety, noting the staircases that connect the arcade to the outside square and worrying that the benches might move too close to the edge and fall down the steps. 

\subsubsection{Gestural timing}
Participants interpreted the intention of a robotic bench based on how the timing of its gestures aligned with their own actions.
Eleven participants framed the robotic gesture as directly connected to their \textit{own presence}, for example, describing the benches as \textit{“adapt[ing] to me”} (F23) or \textit{“mov[ing] because we came in”} (F47). Their interpretations also depended on their own directionality, with the same gesture understood as “coming toward me” (n=4) when facing the bench, or as “follow[ing] me” (F13, F24) when moving in the same direction. F27, who noticed the bench while approaching the library entrance, interpreted its gesture as accommodating their actions, describing it as \textit{“moving away so we could get in.”}

\subsubsection{Robotic familiarity} 
Lastly, participants also interpreted the intention of the benches based on their resemblance to other robots that they were familiar with.
Some (n=5) related it to service robots, describing it as \textit{“one of those delivery robots”} (F10) or likening it to an \textit{“automatic vacuum cleaner”} (F26). F10 also projected typical robotic capabilities onto the bench, suggesting it was \textit{“trying to scan its surroundings”}.

\section{Discussion}






Looking at the different foci of engagement collectively, along people's interpretations of intentions, our results demonstrate the \textit{transitional affordances} of the robotic public benches, in how they shifted seamlessly from manifesting a \textit{robotic affordance}---where passersby perceived them as innovative robots that captured attention and curiosity, to a \textit{spatial affordance}---where passersby perceived them rather as interaction mediators that guided an exploration of the surrounding environment, and ultimately to an \textit{infrastructural affordance}---where passersby perceived them as mundane urban furniture that retreated into parts of the socio-spatial context. \hl{Different from the audience funnel in public display research, which focuses on stages of engagement with the intervention itself~\cite{Muller2010DesignSpace,Michelis2011funnel}, these transitional affordances show how passersby’s engagement shifts as the robotic benches become differently situated within the surrounding space, unfolding} across three phases: first by \textit{activating} engagement through robotic attention, then by \textit{redistributing} that engagement into spatial elements of the surrounding environment, and finally by \textit{settling} into the background and thus allowing engagement to naturally flow into everyday activities. For each phase, we discuss how the three affordances interplayed throughout the gestural sequence of the robotic furniture staged by the socio-spatial context, and compare its impact with prior knowledge from urban HCI and Human-Robot Interaction. We then unpack how these affordances can be actively operationalised to engage passersby in a proactive way.


\subsection{Activating engagement}
This phase describes how robotic public furniture draws the attention of passersby, thereby activating engagement. In this phase, the \textit{robotic affordance} is foregrounded, while its contrast with the {infrastructural affordance} intensifies attention and affords further exploration. 

\subsubsection{Robotic affordance}
Our furniture robots were able to activate the engagement of passersby by leveraging the \textit{robotic affordance} manifested through the presence of their gestural performance alone, which elicited a sense of curiosity, irrespective of interpretation.
Evidently, the gestural performance of our robotic benches drew the attention of most passersby (84.7\%), prompting them to linger in the arcade area where they would otherwise have passed through quickly. At times, passersby even deviated from their intended pathways or returned after leaving the path to further interpret their gestures (see Section~\ref{sec:sittingPACD}).

When set in motion, ordinary objects can be perceived as robotic~\cite{Brown2024Public,Sirkin2015Ottoman}. As autonomous robots are still relatively novel in the urban environment, their presence naturally attracts attention and curiosity from passersby~\cite{Weinberg2023Sharing,Cheon2025Delivering,Pelikan2025MakeSense}, which motivates exploratory behaviours~\cite{Xinyan2024Understanding,Pelikan2024Encountering}. In our case, the \textit{robotic affordance} emerged immediately through the dynamic motion of a bench unit in a similar manner to other mobile urban robots such as delivery~\cite{Xinyan2025Breakable, Pelikan2024Encountering} or service~\cite{Nielsen2022Youtube} robots. On top of this, the relatively large, stable, and seemingly “anchored” physical morphology of the bench unit, which naturally afforded sitting, stood in contrast \hl{to its movement}. This conceptual tension between \textit{sitting} and \textit{moving} thus intensified a sense of \textit{ambiguity} in how the bench was perceived, a quality known to provoke \hl{curiosity} in interactive artefacts~\citet{Gaver2003}, \hl{which} invites attention and encourages spontaneous use~\cite{Brown2024Public}. 
\hl{In addition, the \textit{robotic affordance} prompted passersby to attribute intentionality to and anthropomorphise the benches, leading them to actively interpret their behaviours. Unlike domestic objects whose expressive behaviours foster ongoing relational attribution~\cite{Pradhan2019Box,Cho2025Living}, such attributions in our setting were ephemeral and faded once the bench settled into the environment.}



\subsubsection{Operationalising robotic affordance} 
Whereas the subtle, slow, and quiet deictic gesture of our furniture robots was initially ambiguous to passersby, they were nevertheless able to nudge active engagement as passersby observed, pursued, and performed exploratory behaviours on them (Section~\ref{sec:sittingPACD}), thanks to the appropriate integration of their robotic affordance into the socio-spatial context. 
Considering the framework proposed by \citet{Pelikan2025MakeSense} that explains how people make sense of robotic encounters in public, we argue that this integration was facilitated through three qualities: localism, environment, and sociability. In our case, \textit{localism} was enabled through the physical morphology of the benches---as familiar infrastructure aligned with the character of the site, which led participants to interpret them as legitimate public furniture (Section~\ref{sec:interpretatingcontext}). The \textit{environment} of the quiet, fully pedestrianised library arcade enabled the subtle gestures of the benches to be noticed, whereas they might have been overlooked in a busier context. Finally, \textit{sociability} emerged from place-based expectations, reinforced by the relatively slow speed and ambient gesture of the benches, purposefully designed to complement their morphology, which resonated with the calm atmosphere of the site and inclined visitors to trust their intention and engage with them as part of the public setting. Together, these qualities established the necessary foundation for passersby's subsequent pursuit and further engagement with the benches. \hl{In summary, to operationalise \textit{robotic affordance}, designers should ensure that the motion and morphology of robotic public furniture are carefully aligned with the socio-spatial context, such that the context itself helps bridge the ambiguity of robotic behaviour with the familiarity of urban infrastructure, enabling passersby to trust, interpret, and pursue engagement.}

\subsection{Redistributing engagement} 
This phase describes how the robotic public furniture redistributes passersby’s engagement with itself to the surrounding spatial context, and eventually anchors their attention on an otherwise overlooked element within it. At this stage, the \textit{spatial affordance} comes to the fore, as passersby engage indirectly with the socio-spatial context, mediated by the robotic gesture.

\subsubsection{Spatial affordance}
\label{sec:spatialaffordance}

Through a deictic gesture performed by a single unit that guided passersby's attention across the arcade, and then an orchestrated architectonic gesture involving three units that anchored an existing sculpture as a focal point, our furniture robots collectively enabled their \textit{spatial affordance}, redirecting passersby’s engagement from themselves to the broader socio-spatial context. 

As evident in our study, a common pattern among passersby was to first follow a gesturing bench unit with their gaze and, once it settled, allow their attention to spill into the surrounding environment as they searched for cues to interpret its gestures (Section~\ref{sec:spillover}). This led 20~\((20.4)\%\) passersby to eventually discover the otherwise overlooked sculpture.
People’s vision tends to follow moving objects and disengage once they stop~\cite{Brown2007Attention}, a mechanism widely leveraged to direct attention in digital interfaces through the notion of \textit{Moticons}~\cite{Bartram2003Moticons}. It demonstrates that motion cues not only attract attention in peripheral vision but also guide it toward new interface locations, effectively directing users to where interaction is required. Our robotic benches extend this notion from digital interfaces into the physical world through deictic gestures, where their motion reallocated the attention of passersby between different zones of an architectural space---here playing the role of a physical “interface”, thus creating the preconditions for further engagement to unfold in the new spatial location.
Our study also validated the effectiveness of the architectonic gestural strategy~\cite{Nguyen2025ArchitectonicGestures} in \textit{indexing} spatial features to convey autonomous robotic intention in a public setting. While this strategy was originally proposed for a single furniture robot in indoor, semi-private environments, where familiar occupants already possessed spatial knowledge and thus could promptly notice spatial changes, our findings demonstrate that unfamiliar passersby who were already engaging with the benches could likewise notice, interpret, and even understand the intentions conveyed through the spatial relation between multiple robotic benches and their surroundings.

Using robots to guide spatial exploration is not new. Museum guide robots, for example, have long been deployed to lead visitors through exhibitions, accompanying them around exhibits while employing humanoid referential cues (e.g., pointing gestures, head orientation)~\cite{Pitsch2014Museum} and verbal speech~\cite{Iio2020HumanGuide} to direct attention and contextualise space. However, their very presence can also impose an additional layer on the spatial experience, at times distracting from the exhibition's context. Inspired by the notion of \textit{calm technology} introduced by \citet{weiser1996designing}, we argue that our furniture robots were able to move fluidly between the \textit{center} and \textit{periphery} attention of passersby. By interchanging between deictic and architectonic gestures, they leverage their \textit{spatial affordance} to guide exploration, while their \textit{infrastructural affordance} enables them to settle into the environment and avoid overwhelming or disrupting the overall context.

\subsubsection{Operationalising spatial affordance}
The discovery of the sculpture was not solely the effect of the gestural sequence, but the outcome of a collective interplay between the three units of the robots, passersby, and the surrounding context. This was evident in the diversity in how people interpreted the robotic intention~(Section~\ref{sec:Interpret}), which led to their discovery of the sculpture not only through inferred meaning but also through more incidental encounters~(Section~\ref{sec:spillover}).


Using the lens of \textit{distributed cognition} theory~\cite{Hollan2000DistributedCcognition,hutchins1995cognition}, we argue that the gestural sequence of our robotic benches did not carry intrinsic meaning by itself, but became meaningful through their integration into a distributed cognitive system encompassing \textit{artefacts}, \textit{environment}, and \textit{people}.
Within this system, the two robotic benches acted collectively with the static unit as a mediator that reoriented attention, while the environment provided cues for interpretation, from which passersby constructed meaning that determined their engagement. 
First, the designed gestures of robotic public furniture to effectively guide exploration, as unpacked in Section~\ref{sec:spatialaffordance}. Second, the cultural context of the location as a place that hosts multiple artworks shaped how the intention of the robotic bench was interpreted.
For example, the same installation may attract attention in an expo setting but may be overlooked in a busy shopping street, where it is perceived merely as part of the commercial environment~\cite{Dalsgaard2010mediafaccades}. Third, passersby’s expectations shaped by their role as visitors constructed the benches’ \textit{spatial affordance} within the socio-cultural context of the site, as they entered anticipating encounters with artworks and cultural features.
This phenomenon thus resonated with previous Urban HCI research showing how socio-cultural contexts influence perceptions of interactive artefacts~\cite{Memarovic2013PLAYERS,Dalsgaard2010mediafaccades}. \hl{In summary, to operationalise \textit{spatial affordance}, designers should carefully curate deictic and architectonic gestures so that they fluidly shift between foregrounding specific spatial elements and receding into the background, in ways that remain sensitive to the socio-cultural context of the site.}


\subsection{Settling engagement} 

This phase captures how the robotic public furniture settles into the background, with its \textit{infrastructural affordance} taking the centred stage as a robust, stable \textit{place} to sit, while its \textit{spatial affordance} enables passersby to linger in a meaningful way.

\subsubsection{Infrastructural affordance}
As our furniture robots stopped gesturing and settled into place, their \textit{infrastructural affordance} came to the foreground, fostering familiarity both through their physical morphology and through accumulated engagement over time in previous phases.
This was evident in our study, where five participants naturally chose to rest on the benches after the initial excitement of discovery had faded (Section~\ref{sec:sittingPACD}).

Previous work on robotic furniture has underscored the important role of their \textit{infrastructural affordance} in cultivating familiarity, which in turn, supports trust in their \textit{robotic affordance}~\cite{Nguyen2022Responsive}. In contrast, when \textit{robotic affordance} conflicts with \textit{infrastructural affordance}, this misalignment can disrupt expectations and discourage people from using it. In an in-the-wild study of a shape-changing public bench, unexpected alterations of its form while being occupied were found to startle people and even cause social discomfort~\cite{Kinch2014ShapeChanging}.
Our study instead evidenced a different dynamic, in which passersby still intentionally chose to engage with a robotic bench by sitting on it, even after observing it autonomously move away from them---behaviour that should have contradicted the expectation that a robotic bench would approach people as an invitation to sit. While the motion inevitably introduced a degree of unfamiliarity to our robotic bench, passersby’s willingness to engage with it afterwards can be attributed to three factors: the familiar physical morphology of the bench that afford sitting, its symmetrical spatial configuration with the other two units around the sculpture that signals a sense of stability, and the willingness of passersby in participating in a potential artistic installation. Furthermore, the absence of explicit interaction interfaces also nudged more exploratory passersby to perceive sitting, its functionality embedded in the familiar form, as the primary interaction modality.

\subsubsection{Operationalising infrastructural affordance}
After remaining static for a period of time, our furniture robots gradually blended into the surroundings and was taken up as part of the public infrastructure, which passersby naturally utilised. 
Such blending, however, requires the \textit{infrastructural affordance} of the furniture robots to establish a meaningful connection with the socio-spatial context of its new anchor location, thereby enabling passersby to appropriate it in ways that suit their own needs. 
More than half of the participants who sat on the bench (24 out of 33) chose the side facing away from the sculpture, contrary to our intended orientation. Interviews revealed that participants appropriated the bench according to their own needs, for example, facing the pathway to better observe the flow of people while waiting for friends, or turning outward toward the square to watch events taking place there (see Section~\ref{sec:orientation}). 

\citet{Dix2007appropriation} highlights \hl{that technologies should be designed to support} how people adapt them to their own situated purposes rather than following designers’ intentions. Our robotic benches manifested the guidelines of \textit{designing for appropriation} in several ways. First, they \textit{allowed interpretation} by being settled in ways that made sense within the context, for example, by not obstructing passage, which legitimised them and opened up interpretive possibilities. Their placement offered open views to the pathway and outdoor spaces, further enabling people to project their own meanings onto the configuration and build personal connections to the site. Second, they \textit{supported rather than controlled} use through the open-ended morphology of their \textit{infrastructural affordance}. The absence of armrests, for instance, avoided constraining behaviour and enabled people to appropriate the benches in their own ways. \hl{In summary, to operationalise \textit{infrastructural affordance}, designers should ensure that robotic public furniture settles in ways that legitimise its presence within the socio-spatial context and support open, flexible appropriation by passersby.}

\subsection{An Affordance Transition Model for public robotic furniture}

\begin{figure*}[h]
\begin{center}
\includegraphics[width=0.95\textwidth]{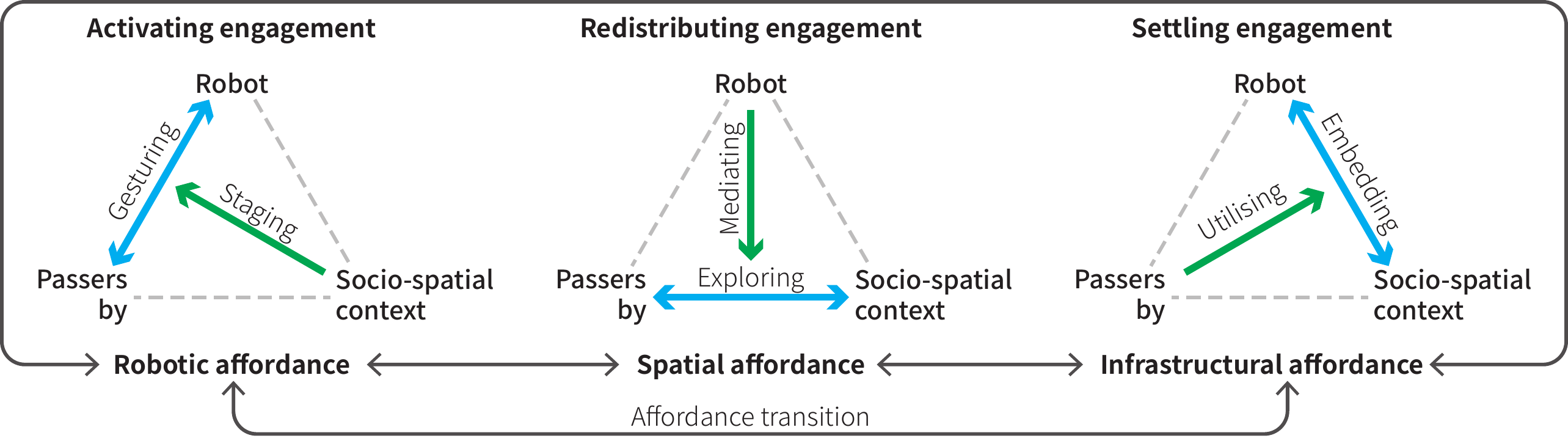}
\end{center}
\vspace{-5pt} %
\caption{Our Affordance Transition Model (ATM) describes the three affordances---robotic, spatial, and infrastructural---that a public furniture robot could manifest, depending on how the interaction unfolds between three main actors: the robot, the passersby, and the socio-spatial context. While the transition between these affordances can be proactively facilitated through the design of the robot gestural sequence, passersby might initiate engagement out of sequence or even enact multiple affordances simultaneously.}
\label{model}
\Description{This figure illustrates the Affordance Transition Model (ATM) for robotic public furniture. The diagram shows three stages of affordances—robotic, spatial, and infrastructural—arranged in sequence along a horizontal axis labeled Affordance transition. In the robotic affordance, a triangle connects three actors: the robot, passersby, and the socio-spatial context. The arrows indicate that the robot interacts with passersby, and the socio-spatial context provides a stage for this interaction. In the spatial affordance, the robot mediates attention between the passersby and the socio-spatial context, positioning itself as a guide. In the infrastructural affordance, the triangle again connects the three actors, but here the socio-spatial context is emphasised: passersby interact with the context, and the robot is shown as part of the environment to be utilised like ordinary furniture.}
\end{figure*}

From our findings, we propose an Affordance Transition Model (ATM) that demonstrates the unique capacity of public robotic furniture to manifest three distinctive affordances for passersby: robotic, spatial, and infrastructural. As illustrated in Fig.~\ref{model}, we argue that these affordances unfold through the interaction of three main actors: the robot, the passersby, and the socio-spatial context. \hl{At the same time, while the PACD model~\cite{Memarovic2012DiscoveryPublicSpace} describes how public displays can stimulate passive engagement, active engagement, and discovery in public space, our ATM model shows that these forms of engagement can be dynamically foregrounded as the robotic furniture transitions between these affordances.} Specifically, the \textit{robotic affordance} emerges when the \hl{robot’s \textit{gesturing} attracts passersby’s attention while the socio-spatial context \textit{stages} the encounter, foregrounding passive engagement as people watch and make sense of the motion.} The \textit{spatial affordance} arises when the \hl{robot’s gestures \textit{mediate} the flow of attention through the surrounding environment, passersby begin \textit{exploring} contextual cues, enabling active engagement and, at times, the discovery of previously unnoticed site features.} Finally, the \textit{infrastructural affordance} manifests when the robot \textit{embeds} itself into the socio-spatial context and passersby \hl{\textit{utilise} it as part of the environment. Such appropriation first manifests as active engagement and then slowly recedes into the ongoing passive engagement as the robotic furniture settles into place.}


Our findings further demonstrate that the transition between these affordances can be proactively facilitated through the gestural sequence performed by the robotic furniture. Drawing on the Laban Effort Framework~\cite{laban1975modern}, often used in human and robotic choreography~\cite{venture2019robot}, we argue that to activate engagement through robotic affordance, robots should emphasise timing in the \textit{time effort}, ensuring their gestures are visible to passersby. To redistribute engagement through \textit{spatial affordance}, they should emphasise directionality in the \textit{space effort}, guiding attention toward relevant spatial elements that can sustain engagement. To settle engagement through infrastructural affordance, they should emphasise anchoring in the \textit{flow effort}, embedding themselves appropriately into the environment so that engagement naturally flows into everyday activity. Across all phases, gestures should convey a sense of reliability, such as through controlled and stable motions, in alignment with the \textit{weight effort}.

Although our designed gestural sequence prescribed a particular order in which affordances unfolded, we observed deviations during the in-the-wild deployment as passersby initiated engagement out of sequence or even enacted multiple affordances simultaneously. For example, some directly used a stationary robotic bench as a resting place, perceiving its \textit{infrastructural affordance} before realising it was robotic. Others interpreted it directly through spatial affordance, treating it as a barrier or installation within the library arcade~(Section~\ref{sec:interpretatingcontext}). Particular occupants combined multiple affordances while engaging with the bench, by sitting playfully on the bench as it moved (\textit{robotic + infrastructural}), or assuming its motion was intended to open the library entrance (\textit{robotic + spatial}).
These observations suggest that while robotic furniture can proactively facilitate affordance transitions in a defined sequence to best engage passersby, they must also anticipate off-script engagements. For instance, if a bench has already been appropriated as infrastructure by unaware passersby, it should not suddenly activate its \textit{robotic affordance}, as this could cause startle or discomfort. Likewise, when multiple affordances are perceived at once, such as a mobile infrastructure robot, its behaviour should adapt to ensure both safety and continued engagement.

Overall, we argue that our Affordance Transition Model could be transferred to inform the design of other robotic furniture in public space. By recognising and proactively utilising the different affordances, while also anticipating diverse real-world engagements, designers can ensure engaging, safe, and contextually appropriate interactions. Moreover, the three affordances we identify may extend beyond robotic furniture, offering a lens for understanding other autonomous robots operating in human-occupied environments, where they may be perceived interchangeably as robots, socio-spatial actors, or even as infrastructure.

\hl{\subsection{Limitations and future directions}}

\hl{First, given that our study was conducted at a single cultural site, future work could investigate settings with different socio-spatial characteristics (e.g., commercial areas, residential public spaces, or transportation nodes), to evaluate how the affordance transitions unfold across varied public rhythms. Second, our findings were drawn from a short-term deployment focused on first-encounter interactions, without capturing how behaviours evolve through long-term habituation. As robotic public furniture becomes a familiar part of the environment, longitudinal studies will be needed to understand how engagement stabilises or transforms over time. Third, while our work focused on robotic public furniture that affords direct bodily engagement, the Affordance Transition Model may also extend to animating public infrastructure that shapes urban experience more ambiently rather than through direct use (e.g., bollards, shade structures, signage). Exploring how our model translates to such infrastructure presents a promising direction for future design research.}

\section{Conclusion}
This paper investigated how robotic public furniture can mediate social and spatial engagement in public space. Through an iterative design process and in-the-wild deployment of a set of robotic benches, we demonstrated how robotic public furniture can activate passersby's engagement and motivate exploration, redistribute attention toward the surrounding spatial context, and ultimately settle as part of everyday urban furniture. Building on these findings, we introduced an \textit{Affordance Transition Model} that conceptualises how transitions across three affordances enable public robotic furniture to mediate passersby’s engagement with space: a \textit{robotic affordance}, where they were perceived as novel robots capturing attention and curiosity; a \textit{spatial affordance}, where they guided exploration of the surrounding environment; and an \textit{infrastructural affordance}, where they receded into the socio-spatial context as mundane urban furniture. By bridging research on robotic furniture and Urban HCI, our work offers a lens for understanding how familiar public infrastructures can be reimagined as interactive artefacts that activate public engagement and inform the design of future robotic public furniture.



\bibliographystyle{ACM-Reference-Format}
\bibliography{sample-base}


\end{document}